# Automated fabrication technique of gold tips for use in point-contact spectroscopy

S. Narasiwodeyar, M. Dwyer, M. Liu, W. K. Park[*], and L. H. Greene

*Department of Physics and the Frederick Seitz Material Research Laboratory, University of Illinois at Urbana-Champaign, Urbana, Illinois 61801, USA*

Abstract

For a successful point-contact spectroscopy (PCS) measurement, metallic tips of proper shape and smoothness are essential to ensure the ballistic nature of a point-contact junction. Until recently, the fabrication of Au tips suitable for use in point-contact spectroscopy has remained more of an art involving a trial and error method rather than an automated scientific process. To address these issues, we have developed a technique with which one can prepare high quality Au tips reproducibly and systematically. It involves an electronic control of the driving voltages used for an electrochemical etching of a gold wire in an HCl-glycerol mixture or an HCl solution. We find that a stopping current, below which the circuit is set to shut off, is a single very important parameter to produce an Au tip of desired shape. We present detailed descriptions for a two-step etching process for Au tips and also test results from PCS measurements using them.

[*]Corresponding author: wkpark@illinois.edu

## I. Introduction

Point-contact spectroscopy (PCS) is a simple and versatile technique to probe the bulk spectroscopic properties of metals and superconductors[1] and, most recently, has been found to be a probe in correlated electron systems.[2-5] The fabrication of gold tips of proper dimensions and surface quality is vital to a successful measurement with PCS. PCS requires making a stable and small contact on the sample surface with a clean Au tip to obtain conductance spectra. This contact has to form a *ballistic* junction, whose dimensions are smaller than the electronic mean free paths in a given material.[2] In order to fulfill these requirements, the finished Au tip is bent prior to mounting onto a tip holder.[6] Hence, the lower body of the tip, which will be in contact with the sample, needs to have smooth surface and uniformly decreasing diameter.

Electrochemical etching methods of Au tips for use in scanning tunneling spectroscopy, Raman spectroscopy, and near field optical microscopy are well known.[7-16] However, most of these methods focus only on producing a tip of small apex radius, rendering them unsuitable for PCS. Their methods employ the use of AC or DC[12] power to etch a gold wire in a solution usually containing HCl or KCl.[14] Although several of these methods have been established over the years and the electrochemistry is well known,[8,9,17,18] producing a tip of required specifications remains more or less an art skill, requiring a substantial amount of practice. In order to develop an automated procedure, detailed understanding of each of the etching steps along with the parameters[9] and careful data analysis are needed.

Eligal *et al.*[12] report an automated technique that uses DC voltage to etch an Au wire in a solution of HCl and demineralized water. In their paper, they describe a method in which they record the current in the circuit during etching and explore a range of current values where etching can be automatically terminated as the current drops. However, it is lacking data analysis in full details, leaving it unclear what conclusions can be drawn about the possible role and significance of the etching current. On the other hand, there is not much literature regarding the effects of fluid dynamics of the etching solution and their impact on the resultant tip size and quality. Park *et al.*[15] describe an etching method where the formation of a meniscus[10,15] plays a key role in the shape and size of the final tip. Still, this is not an automated technique. Thus, it does not provide a full control of the tip etching process. We adopt a part of their technique, make further analysis on the fluid behavior during the etching process, and study the role of the meniscus and its effects on the resulting tip.

In this paper, we present an electrochemical etching technique to fabricate tips specifically suitable for PCS, yielding tips as per the above mentioned requirements for PCS. A two-step procedure has been developed using an HCl-glycerol mixture. Our electrical circuit combines the use of AC and DC powers. For further precise control, the etching circuit is operated using an automated LabView program. The tip's required dimensions are achieved with the first step. Then, the second step of the procedure improves the surface quality of the tip's lower body. The first step is automated by terminating the driving electrical powers when the current in the circuit falls below a fixed value during the first step, and the second step is automated by shutting off the driving electrical powers after a set amount of time. Current is constantly monitored on the computer screen using the LabView program. Alhough the resulting tips are suitable for PCS, as evidenced by our recent PCS measurements using them,[3,19,20] the methods described in this paper can also be used to fabricate tips for various other experimental uses by adjusting the parameters of etching. We also provide an analysis of the effects of various controllable parameters in our setup which can be used to tailor this setup for other uses.

## II. Methods

### A. Apparatus and Setup

To etch a gold wire down to fine tips suitable for PCS, a function generator (Agilent 33210A) and a DC power supply (HP 6227B Dual DC) are connected in series with a 10 ohm high-wattage resistor as shown in Fig. 1. The circuit is completed via a gold wire and a platinum wire as positive and negative electrode, respectively, immersed in a mixture of HCl and glycerol. The current flowing in this circuit is detected by measuring the voltage across the resistor using a DAQ (NI-USB-6210, sampling rate of 250 kS/s). Output parameters of the function generator are controlled through a LabView program. The gold wire is of 0.5 mm diameter (Alpha Aesar, purity 99.995%) and the platinum wire used as a counter-electrode is of 0.2 mm diameter. To prevent the bubbles that form at the Pt electrode from perturbing the solution surface and breaking the meniscus that forms on the gold wire, the Pt wire is put into a small glass tube before being submerged in the solution. The gold wire is straightened and inserted through a pipette into the solution to keep the wire still during the etching process.

Two solutions are used during the entire etching process. The first is a mixture of fuming HCl (~175 ml at 36.45%) and glycerol (~12 ml at 99%). The glycerol is added to increase the viscosity of the solution, which helps to form a stable meniscus and etch the gold wire to a fine tip. The second solution is the concentrated stock of HCl (175 ml at 36.45%) without any additions.

## B. Etching Process

The etching process is split into two parts: the first part with the HCl-glycerol mixture and the second part with the pure HCl solution. Various etching equations for Au in HCl with respective potentials are known as follows:[8,9,17,18]

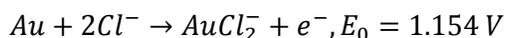
$$Au + 2Cl^- \rightarrow AuCl_2^- + e^-, E_0 = 1.154\,V$$

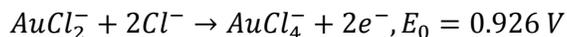
$$AuCl_2^- + 2Cl^- \rightarrow AuCl_4^- + 2e^-, E_0 = 0.926\,V$$

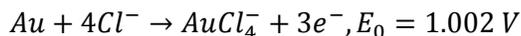
$$Au + 4Cl^- \rightarrow AuCl_4^- + 3e^-, E_0 = 1.002\,V$$

To begin the etching process, the Pt electrode surrounded in its glass tube is immersed into the solution approximately by 10 mm, and approximately 1 mm of the gold wire is dipped into the solution. The exact length of the wire is irrelevant so long as there is enough portion of the wire submersed in the solution for adequate current flow. The distance between the wires is kept roughly constant at 35 mm. As the gold wire is dipped into the solution, a meniscus forms around the wire due to the surface tension of the fluid.[10] Once the wire is submerged an acceptable amount and is as perpendicular to the solution as can be, a combined voltage from the DC power supply and the function generator is applied through the LabView program. Etching begins immediately and the current is being recorded and monitored. Power to the circuit is shut off automatically through LabView when a certain current is reached. Hereafter, we refer this current to as a stopping current, $I_{stop}$. Optimal electrical parameters found are as follows: a DC offset of total 6 V (3.5V from the function generator plus 2.5 V from the DC power supply), an AC pulse of 1 kHz (3 V peak to peak amplitude, 40% duty cycle), and $I_{stop}$ = 54 mA, so the voltage applied to the circuit alternates between 6 V and 9 V throughout the etching process.

The current oscillates around a fixed value before slowly decreasing over time as shown in Fig. 2. Violent and constant bubbling of $H_2$ gas is observed around the Pt electrode, validating the necessity of the glass tube encasing. Less intense bubbling of $Cl_2$ gas is also observed at the gold wire, but not so intense as to disrupt the meniscus formed.

Etching occurs in two stages. The first stage accounts for the majority of the process and involves the immersed wire being etched away from the outer edge of the wire inward. This etching leaves a long and thin shaped tip under the surface (Fig. 3b). Once this reaches a certain point, the final stage of etching beings. Since the remaining gold wire is volumetrically small and therefore the current density is high around the wire, the remaining part is etched away very quickly, rapidly diminishing its vertical length (Fig. 3c). This corresponds to the fast drop in the current near the end of the etching process. However, etching does not stop immediately because the surface contact is still maintained with the meniscus on the upper part which is not immersed as it is above the solution surface (Fig. 3c & 3d). During this last phase, the formation of the meniscus plays a key role in the shape of the tip. As the current value reaches the stopping current, the power is automatically cut off via the LabView program. The choosing of the stopping current is paramount to the final tip shape. An ideal stopping current is well below the initial current (~100 mA at the setting described above) but not so low that the tip keeps etching after an ideal shape has been reached. This stopping current can be found by observing the current vs. time graph and choosing a point after the dropping point (Fig. 2). Inspection of the tip shows a long and thin shape with a uniformly decreasing radius (Fig. 4). Because of the influence of the meniscus, a rougher surface develops towards the lower body of the tip. Total etching duration ranges between 1 and 2 minutes, increasing with every subsequent etch using the same solution. It is found that the same solution is effective and reusable for up to five runs.

The second etching process is used to eliminate the rougher surface on the lower body of the tip caused by the meniscus that forms from the addition of glycerol into the solution. This smoothing process is done in pure HCl. The tip is re-immersed into the bulk solution after going through the first process, and a constant voltage of 6 V is supplied for 100 ms. The bulk HCl doesn't form as strong a meniscus as the HCl-glycerol mixture, so the effects of the meniscus are negligible in this second process. The tip is then removed and cleaned with unheated de-ionized water and methanol. Here, the duration of current supply is once again controlled by the LabView program. This step improves the surface quality of the lower body of the tip, but causes a slight decrease in length. The optimal duration lies between 50 ms and 200 ms depending on the shape and size of the tip and level of roughness. Etching for durations longer than 200ms at this voltage is found to blunt the tip.

## III. RESULTS AND DISCUSSION

For the first step in the etching process (HCl/glycerol), it is found that a combination of low (3 – 7 V) and high voltages (> 7 V) gives the optimal results. A DC offset of 6 V was used in parallel with an AC pulse of 3 V peak to peak amplitude. With an AC pulse of 1 kHz, the duty cycle can also be manipulated easily to optimize the etching process. Frequencies of around 1 kHz are found to provide the best results. $I_{stop}$ of 54 mA gives the optimal results: not so much etching occurs to cause the tip to be short and blunt, while not too little etching occurs to produce a tip too thick to be used for PCS. For the second etching step (HCl only), a DC offset of 6 V without any AC pulse is found to give the smoothest lower body on the tip, although this is highly dependent on the geometry of a tip including the length and the cone angle. It should also be noted that the second etching is not always necessary: if a tip concludes the first etching process and is smooth and sharp, then the second etching could blunt the tip.

**A. Effect of Applied Voltage and Frequency**

The first step etching process proceeds uniformly and rapidly at voltages of above 7 V but a rapid bubbling occurs at the gold wire. This causes the etching to be non-uniform, affecting the surface quality of the tip. At a lower voltage, namely, between 3 V and 7 V, the bubbling is less intense but the voltage is not high enough to drive the reaction smoothly and uniformly. AuCl$^-$ salts adhere to the surface of the tip, affecting the quality of the tip surface (cleanliness and smoothness).

Because of these issues with both high and low voltages, a combination of high and low voltages is used within a wave period as mentioned previously. The duty cycle of an AC pulse of 1 kHz can be adjusted easily to optimize the etching. With a duty cycle greater than 60%, resulting tips are sharper and rougher. With a lower duty cycle less than 35%, smoother and blunter tips are produced. Several different frequencies are tested, and it is found that frequencies above 10 kHz result in non-uniform tip shape and rougher surface, whereas frequencies below 500 Hz produce blunt tips. For our PCS needs, frequencies around 1 kHz are optimal. For the second etching step, a constant voltage is optimal due to its short duration. An applied voltage of 6 V gives the smoothest lower body surface on the tip.

**B. Current vs. Time**

The current oscillates throughout the etching as shown in Fig. 2, which occurs due to the production of $AuCl_4^-$ salts from the electrochemical reactions listed in the equations previously. The last two reactions shown in those equations produce $AuCl_4^-$ salts, which naturally adhere to the Au (+) electrode, blocking the current flow. However, as more of these salts are accumulated around the electrode, they drop away in clusters. Because of this, the current oscillates until the etching process finishes.

There are two phases of etching in the first etching process. The boundary between these two phases is seen as a jump in the current as seen in Fig. 2. In phase A, the current oscillates rapidly because of the formation and deposition of salt clusters. In phase B, the current starts to decrease rapidly because of an increase in resistance which can be attributed to a decrease in contact area of the gold wire within the solution (Fig. 3b → 3c). A decreasing current correlates with the diminishing size of a tip within the solution.

When the gold wire is immersed into the solution, a meniscus forms around the wire near the solution surface due to the surface tension of the HCl-glycerol mixture.[10,15] The etch rate for a constant voltage will be dependent on the abundance of $Cl^-$ ions near the surface and, because of this, the gold wire is etched more quickly in the bulk of the solution than near the surface.

After the jump in current (Fig. 2), only a part of the wire above the surface will remain, which will be underneath the meniscus as shown in Fig. 3d. The wire will diminish in size as the etching process continues, allowing the changing meniscus to mold a clean conical shape of the wire (Fig. 3). The meniscus needs to be held under changing conditions without rupturing, so the glycerol is added to the solution to increase its viscosity and surface tension.

At the last stage in Fig. 3d, the current density and, consequently, the reaction rate is quite high. At the meniscus, $Cl^-$ ions are quickly depleted. Because of the lack of $Cl^-$ ions, etching becomes non-uniform and a rough surface texture develops below the meniscus line in the final few moments of etching. This rough surface can be eliminated by re-etching in the second step as described above.

## C. Testing Tips in PCS

Using the Au tips fabricated with the newly-developed method described in this paper, PCS data are taken on a thin Nb film of 2000 Å thickness. The obtained data are analyzed using the Blonder-Tinkhm-Klapwijk (BTK) model.[21] Representative data are plotted in Fig. 5 along with best fit curve to the BTK model. As shown, the conductance data are of Andreev-reflection type with dimensionless barrier strength Z = 0.66. The obtained superconducting gap energy $\Delta$ = 1.49 meV is consistent with the literature value.[6] Also, the small quasiparticle smearing factor $\Gamma$ = 0.29 meV indicates that the point-contact junction is quite clean. This shows that the Au tips prepared using the technique developed in this work are suitable for PCS as desired.[3,19,20]

## IV. SUMMARY

An automated technique for the fabrication of Au tips has been developed. Employing an electronic control of the two-step electrochemical etching process, this method produces Au tips suitable for use in PCS. With given DC and AC voltages, an optimal termination current in the first step using a mixture solution of HCl and glycerol is found to be 54 mA. We refer this to as a stopping current, a single very important parameter to produce a tip of desired shape. A larger stopping current leads to a thicker tip, whereas a smaller stopping current gives a shorter and blunt tip. If necessary, the second etching step lasting for 100 ms in fuming HCl improves the surface quality substantially. Further improvement of the tip quality could be achieved by annealing the gold wire prior to etching.[7,13] Also, use of a smaller diameter gold wire could reduce reaction times and produce a thinner tip. As shown by our PCS measurement on Nb, the Au tips produced by this automated system enable us to obtain high quality data, indicating that they are suitable for PCS.

## ACKNOWLEDGMENTS

We thank H. Zhao for his help with SEM measurements. This material is based upon the work supported by the U.S. NSF DMR under Award 12-06766.

**FIGURE CAPTIONS**

Figure 1. Schematic diagram of the setup used for Au tip fabrication. The solution used is either stock HCl or a HCl-glycerol mixture. The current is monitored by measuring the voltage across the resistor. The etching circuit is controlled by a LabView program.

Figure 2. Current vs. time during the etching of a tip in an HCl-glycerol mixture. The inset shows a zoomed-in portion of the graph to illustrate the oscillations in detail. This graph was recorded when the stopping current was set to 36 mA, instead of the optimal value, but still shows the two phases of the etching process and the oscillations clearly as indicated.

Figure 3. Schematic cartoon pictures to illustrate the effect the meniscus plays in the formation of an Au tip. The time between successive steps is largely dependent on the amount of gold wire initially immersed in part (a) and also on how many times the same solution has been used for the etching process.

Figure 4. Scanning electron microscope image of a processed Au tip. Note the relatively uniform tapering off to a fine point.

Figure 5. Differential conductance data (open circles) taken at 4.38 K from Nb thin film using a Au tip. The solid line is a fit curve to the BTK model.

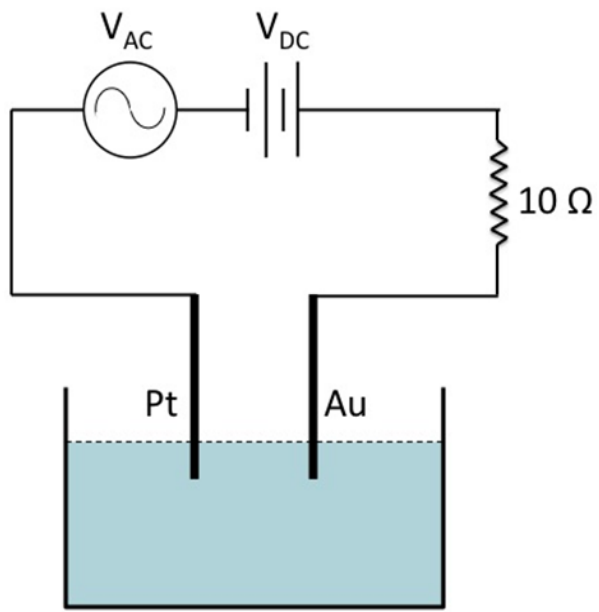

Figure 1

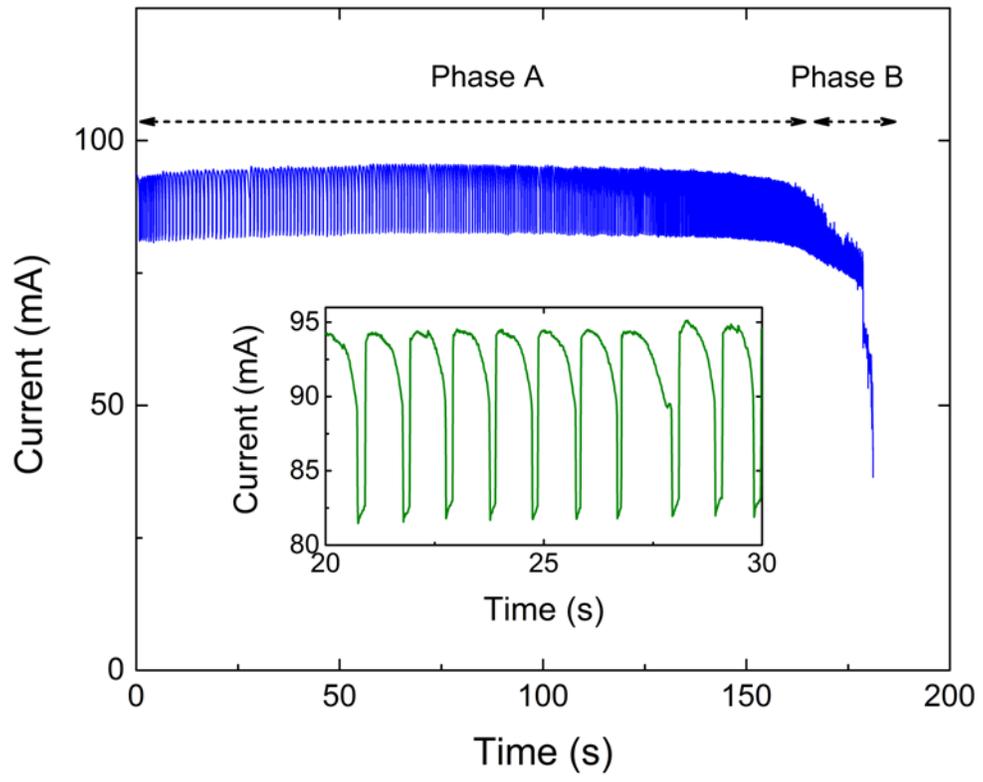

Figure 2

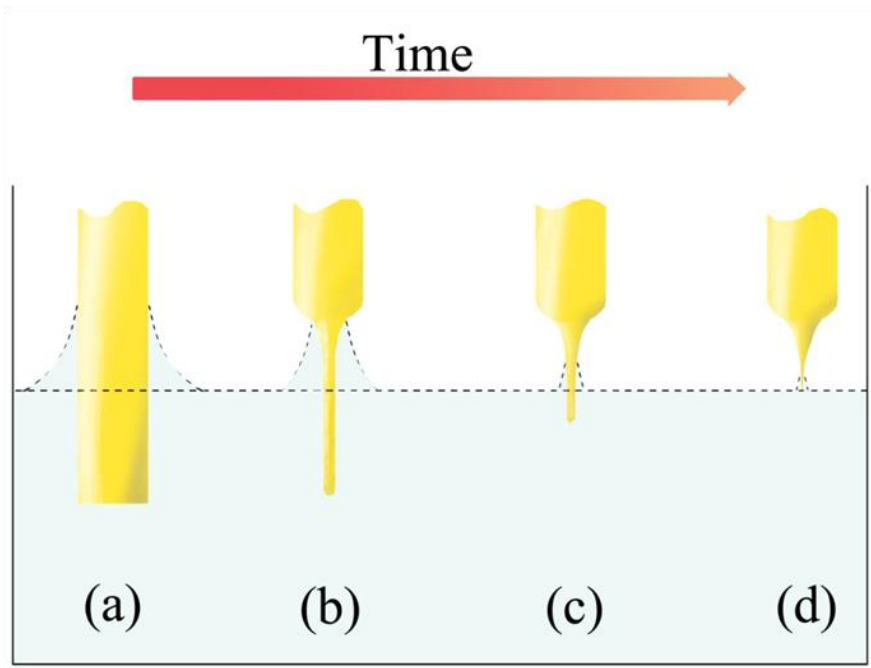

Figure 3

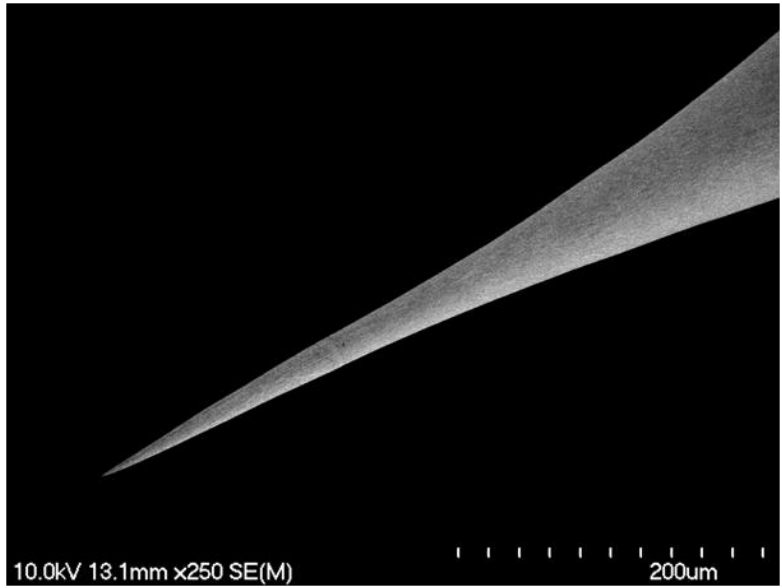

Figure 4

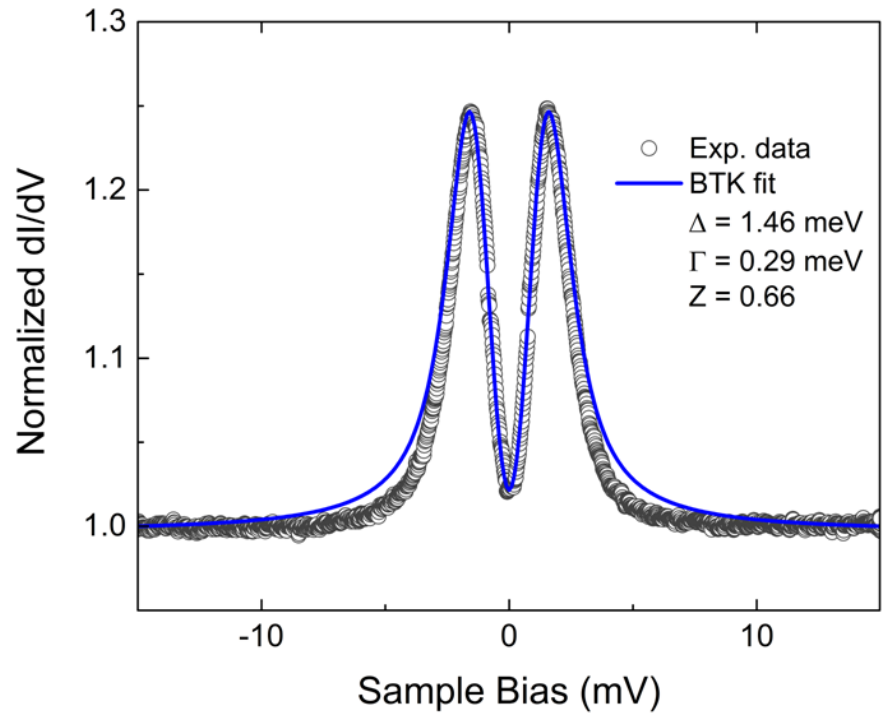

Figure 5